\documentclass[journal abbreviation, manuscript]{copernicus}
\usepackage{epsfig,amsmath,amsfonts,amssymb,graphicx,subfigure,enumitem,color}
\usepackage{amssymb}
\usepackage{amsmath}
\usepackage[varg]{txfonts}
\usepackage{natbib}
\usepackage{url}             
\usepackage{graphicx}
\usepackage[percent]{overpic}
\usepackage{mathptmx}
\usepackage{anyfontsize}
\usepackage{t1enc}
\usepackage{url}
\usepackage{float}
\usepackage{xcolor}
\graphicspath{{./}{figures/}}

\newcommand{\km}{{\ \mathrm {km}} }

\newcommand{\s}{{\ \mathrm s} }
\def\arcsec{\hbox{$^{\prime\prime}$}}

\begin{document}

\title{Signatures of red-shifted footpoints in the quiescent coronal loop system}


\Author[1]{Yamini K. Rao}{}
\Author[1]{Abhishek K. Srivastava}{}
\Author[2]{Pradeep Kayshap}{}
\Author[1]{Bhola N. Dwivedi}{}

\affil[1]{Department of Physics, Indian Institute of Technology (BHU), Varanasi-221005, India.}
\affil[2]{Institute of Physics, University of South Bohemia, Brani\v sovsk\'a 1760, CZ -- 370 05 \v{C}esk\'e Bud\v{e}jovice, Czech Republic.}


\runningtitle{Doppler shifts in quiescent coronal loops}
\runningauthor{Yamini K. Rao}
\correspondence{Abhishek K. Srivastava (asrivastava.app@itbhu.ac.in)}


\firstpage{1}
\maketitle
\begin{abstract}

We observed quiescent coronal loops using multi-wavelength observations from the Atmospheric Imaging Assembly (AIA) onboard the Solar Dynamics Observatory (SDO) on 2016 April 13. The flows at the footpoints of such loop systems are studied using spectral data from Interface Region Imaging Spectrograph (IRIS). The Doppler velocity distributions at the footpoints lying in the moss region show the negligible or small flows at Ni\,{\sc I}, Mg\,{\sc II}\,k3 and C\,{\sc II} line corresponding to upper photospheric and chromospheric emissions. Significant red-shifts (downflows) ranging from ($1~{\rm to}~7) \km\s^{-1}$ are observed at Si\,{\sc IV} (1393.78~\AA; $log(T/K)$ = 4.8) which is found to be consistent with the existing results regarding dynamical loop systems and moss regions. Such downflows agree well with the impulsive heating mechanism reported earlier.

\end{abstract}


\introduction  

The active regions dominated by various loop structures are of 
significant importance for the study of coronal heating since 
these loop systems act as a fundamental unit of the solar corona
(\citealt{2006SoPh..234...41K}; \citealt{2014LRSP...11....4R}; \citealt{2015RSPTA.37340269D}). 
Moss is generally associated with plage regions around the active regions sites (\citealt{1999ApJ...520L.135F})
and transition region emission of hot core loops
will provide us with a better understanding of the flows and thus energy transfer mechanism 
between the transition region (TR) and corona.


\begin{figure*}
\begin{center}
\includegraphics[trim=2.5cm 0.5cm 0.2cm 0.2cm,scale=0.6,angle=90]{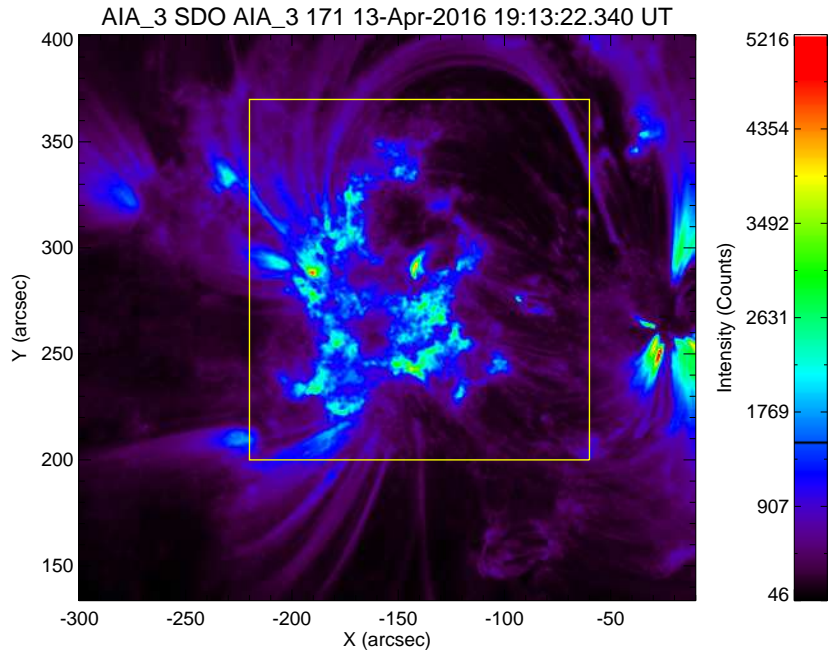}
\vspace{1cm}
\caption{Intensity emission due to 171~\AA wavelength of SDO/AIA at 19:13:22 UT. The yellow box is overlaid to show the region of interest (ROI) taken to analyse the flows at the footpoints of quiescent coronal loops.}
\label{fig0}
\end{center}
\end{figure*}



\begin{figure*}
\begin{center}
\includegraphics[trim=2.5cm 0.5cm 0.2cm 0.2cm,scale=0.6,angle=90]{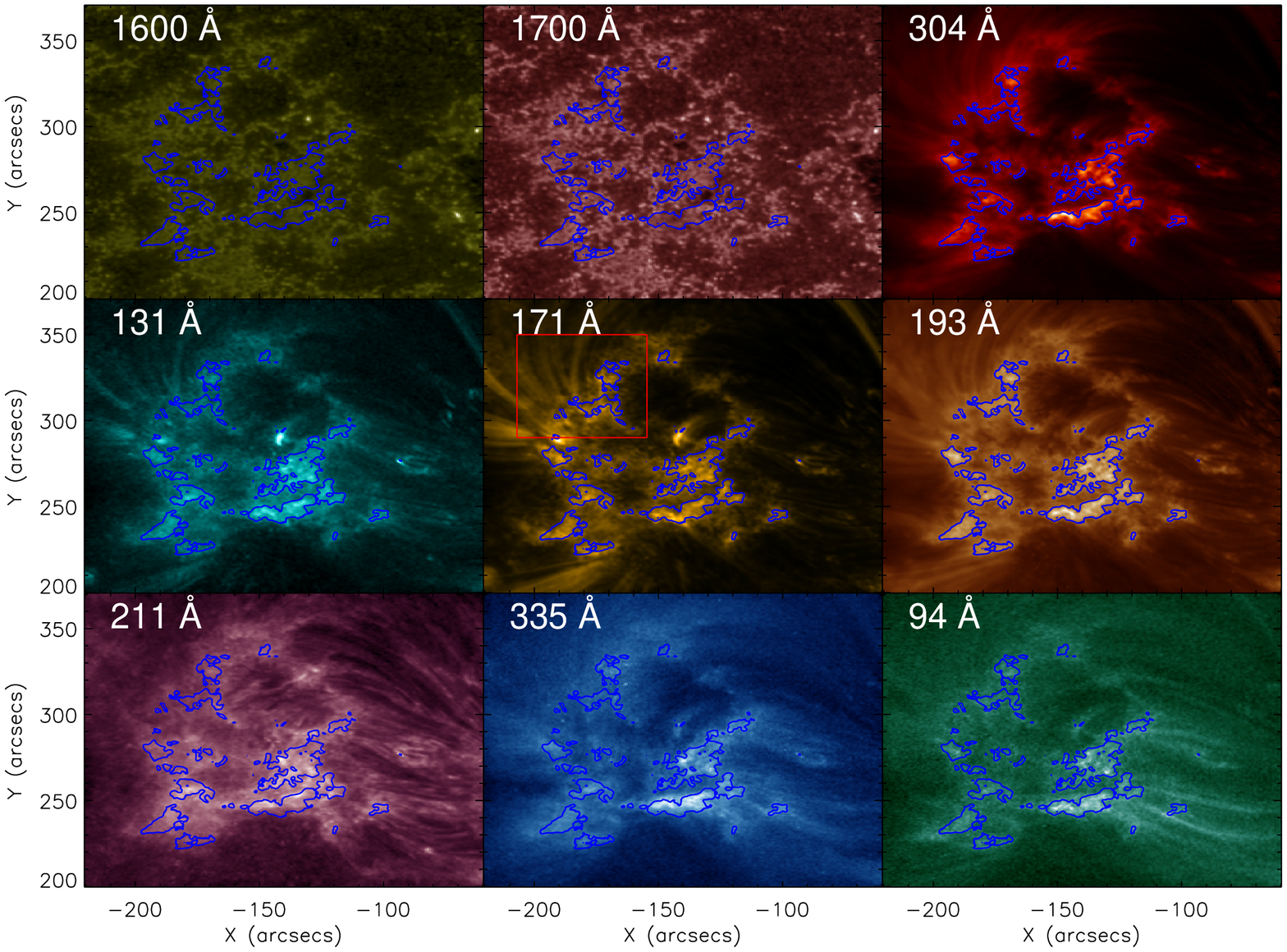}
\vspace{1cm}
\caption{Mosaic representation of the zoom-in-view of the region of interest at different wavelength of SDO/AIA as mentioned on the corresponding panels.}
\label{fig1}
\end{center}
\end{figure*}


\citet{2006SoPh..234...41K} has provided with a full review of the coronal heating problem. It describes that the coronal heating mechanisms are
impulsive when explored from the perspective of elemental
magnetic flux strands.
It has also been well established that the loop structures emit significantly in the 
solar corona which has been classified depending on their temperatures.
The spectral studies of these loop systems in response to the Doppler shift 
provides a clue to distinguishing between the steady and impulsive heating mechanism (\citealt{2008A&A...481L..49D}; \citealt{2011ApJ...730...85B}).

Various types of loops are 
hot core loops \citep{2008A&A...481L..49D}, 
warm loops \citep{2011A&A...535A..46D}, 
fan loops (\citealt{2012ApJ...744...14Y}; \citealt{2009ApJ...700..762W} and references therein)
and cool loops (\citealt*{2015ApJ...810...46H}; \citealt{2019ApJ...874...56R}) present in the different regions of the solar atmosphere.
The temperature and density diagnostics of quiescent coronal loops have been fairly studied earlier (\citealt{2003A&A...406.1089D}).
However, there have been not many observations regarding the 
flows in the resolved strands/flux tubes of such loops in the solar corona.

In this paper, we study quiescent coronal loops with big loop arches having one of their footpoints 
anchored at the edges of moss region. The different strands in such
large loop systems have been identified using high$-$resolution observations of SDO/AIA
and studied the flows in it mapping the footpoints to the lower region of the solar atmosphere.
Section 2 describes the observational data and its analyses
presenting the details of the data used for our analyses. In Section 3, the results have been reported with their interpretations.
In the last section, the discussions and conclusions are summarized.


\begin{figure*}
\includegraphics[trim=2.5cm 0.5cm 0.2cm 0.2cm,scale=0.6,angle=90]{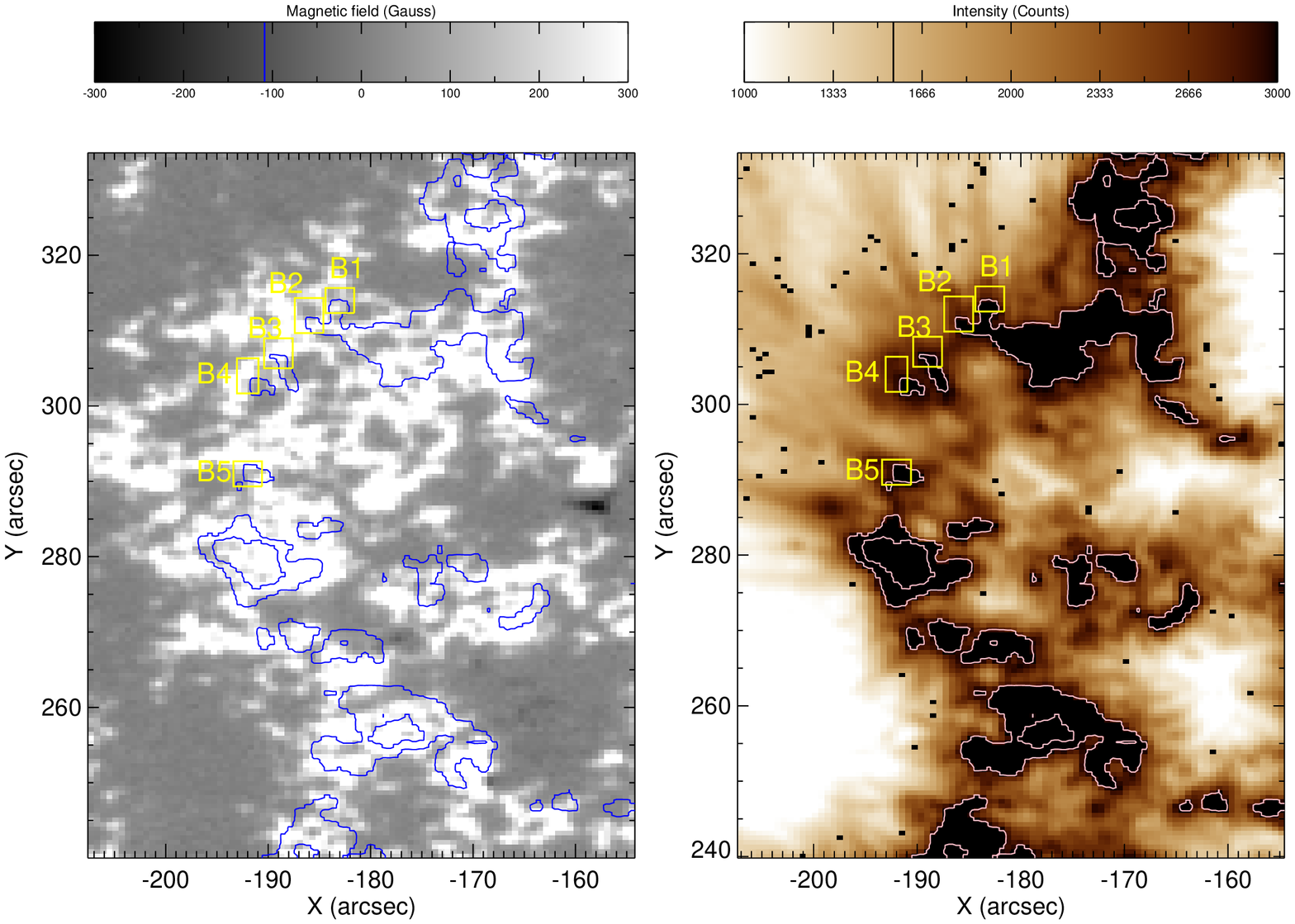}
\vspace{1cm}
\caption{Left panel: HMI map indicating the magnetic polarities at the moss region indicted by blue contours. Right panel: Identification of the footpoints of quiescent loops anchored at the moss regions. The different small boxes are taken at the footpoints of the individual loop strands are shown in both the panels.}
\label{fig2}
\end{figure*}



\begin{figure*}
\includegraphics[trim=2.5cm 0.5cm 0.2cm 0.2cm,scale=0.6,angle=90]{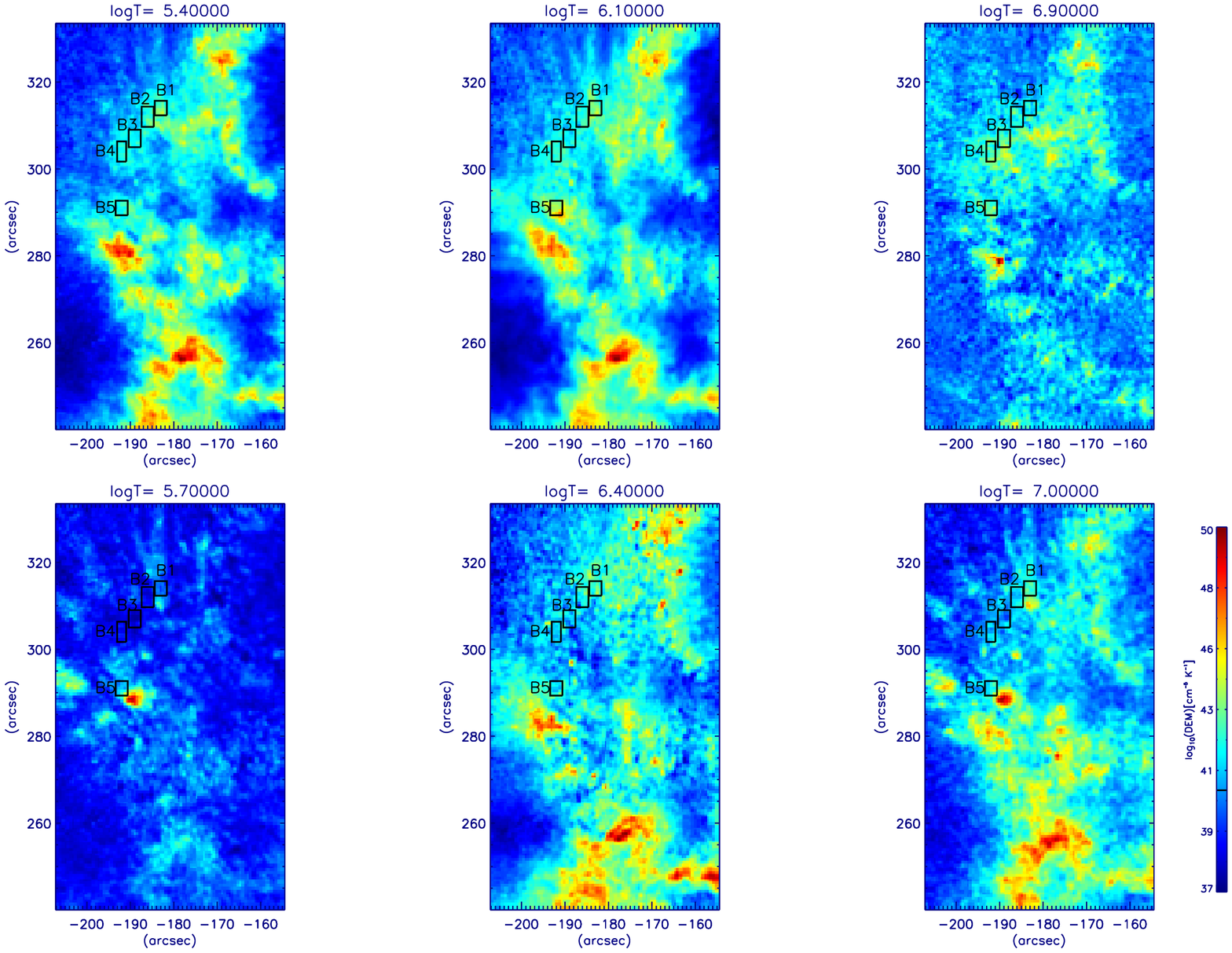}
\vspace{1cm}
\caption{Differential Emission Measure maps of the plage region contaning the moss associated to footoints of quiescent coronal loop systems.}
\label{fig2b}
\end{figure*}

\section{Observational Data}

IRIS provides spectral data in the two UV domains: FUV band (1331.7~\AA~to 1358.4~\AA~and 1389.0~\AA~to 1407.0~\AA) 
and NUV (2782.7~\AA~to 2835.1~\AA)
having a large number of spectral lines covering the photosphere, chromosphere, TR, and inner corona. 
Level~2 data is used for our study, which is calibrated for the dark current 
removal, flat fielding (\citealt*{2014SoPh..289.2733D}).
We have utilized 
Si\,{\sc IV} (1393.78~\AA), 
Mg\,{\sc II}\,k (2796.20~\AA), 
C\,{\sc II} (1334.53~\AA), and 
Ni\,{\sc I} (2799.47~\AA) spectral lines.


\begin{figure*}
\includegraphics[trim=2.5cm 0.5cm 0.2cm 0.2cm,scale=0.6,angle=90]{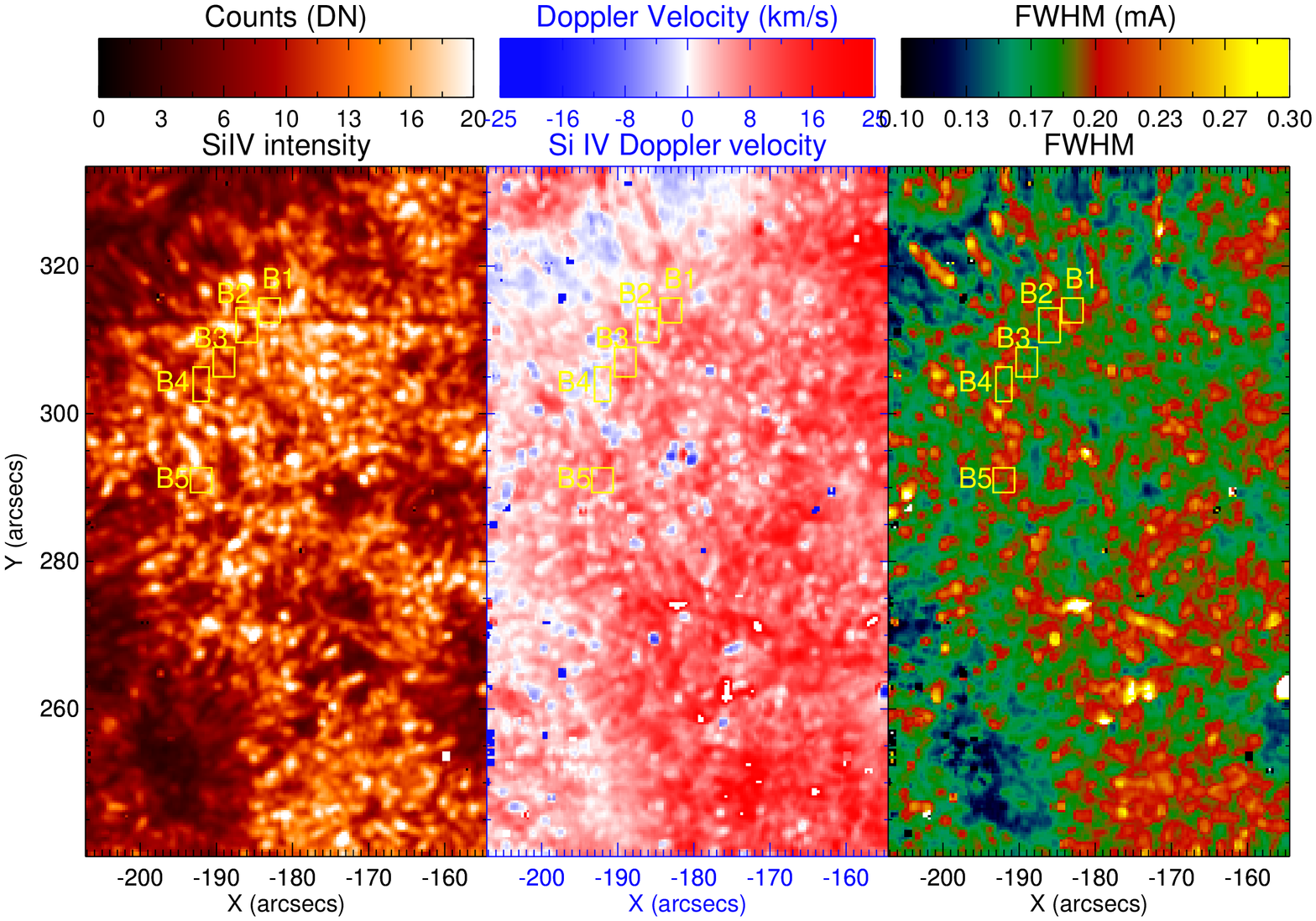}
\vspace{1cm}
\caption{The different parametric plots of  Si IV (1393.78~\AA) line with the footpoints of the quiescent coronal loop systems indicated by different boxes.}
\label{fig2c}
\end{figure*}



\begin{figure*}
\includegraphics[trim=2.5cm 0.5cm 0.2cm 0.2cm,scale=0.6,angle=90]{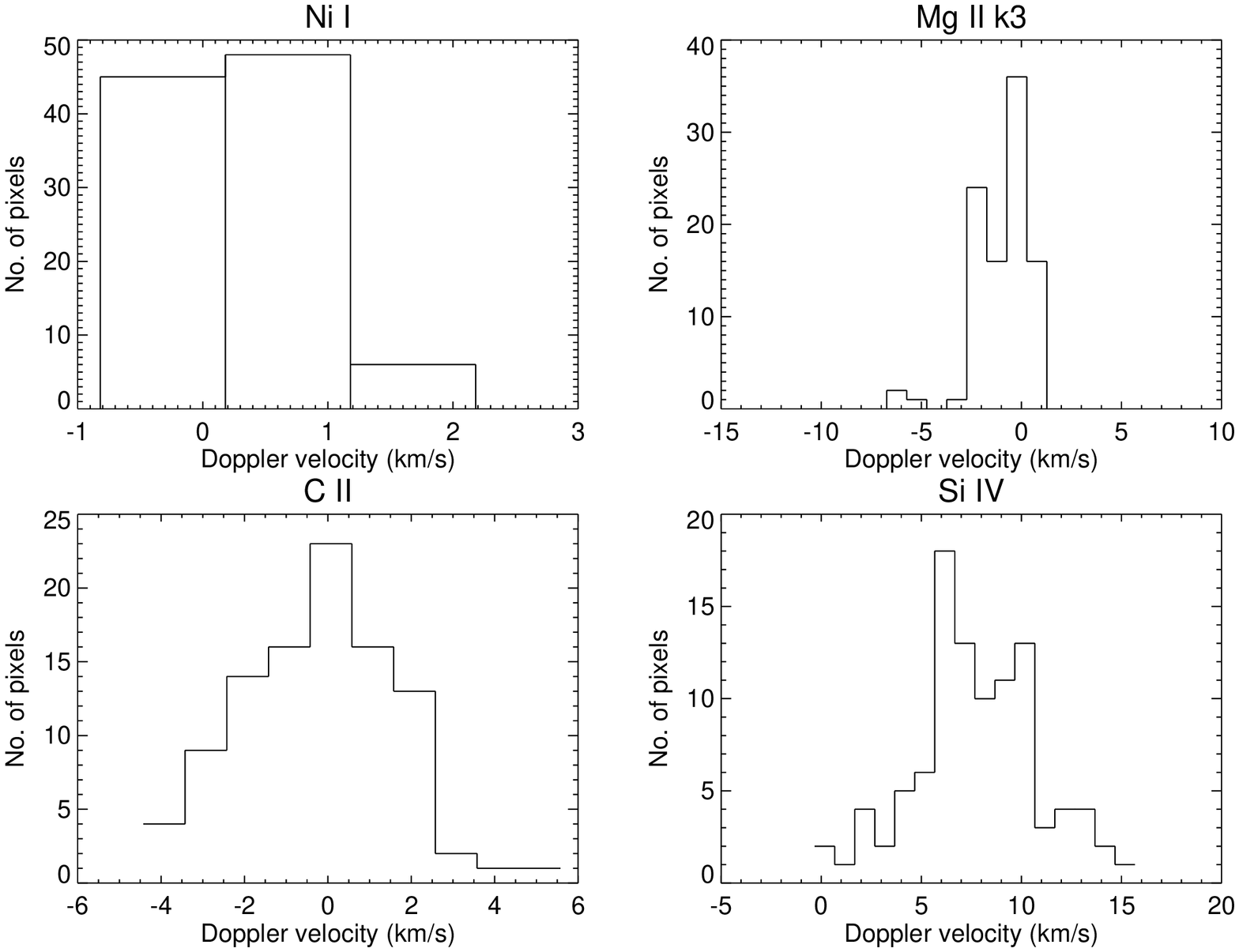}
\vspace{1cm}
\caption{The velocity distributions for different spectral lines corresponding to different temperatures at box B1.}
\label{fig3}
\end{figure*}


In our present work, we have used dense raster data from IRIS for the time period
19:19:09 to 20:21:14 UTC on 2016 April 13
targeting 
the evolution of AR 12529 having slit width of 0.35$\arcsec$ having step cadence of 
9.3 s covering the field-of-view of $141\arcsec$ in x-direction and $175\arcsec$ in y-direction 
centered at the coordinates ($X_{\rm cen},Y_{\rm cen}) = (-173$\arcsec$,275$\arcsec$)$. 
The data is compensated for oscillations due to 
thermal variation using iris$\_$orbitvarr$\_$corr$\_$l2.pro
in the SSWIDL library.
The rest wavelengths for different spectral lines 
used in our analysis are calibrated using 
neutral lines from the relatively quiet-Sun area of the raster. 
The rest wavelength of Ni\,{\sc I} used is 2944.4697~\AA. Mg\,{\sc II}\,k has been calibrated with respect to Ni\,{\sc I} which is found to be 2796.3574~\AA. Si\,{\sc IV} line is calibrated w.r.t Fe\,{\sc I} (1392.8052~\AA) line and C\,{\sc II} is calibrated w.r.t O\,{\sc I} (1355.5987~\AA). So, the calibrated wavelengths used for our analysis is 1393.7604~\AA~and 1334.5406~\AA~for Si\,{\sc IV} and C\,{\sc II} respectively. 

The Doppler velocities are deduced using different spectral lines, i.e., Ni I 2799.47 A, Mg II k 2796.20 A, C II 1334.53 A, and Si IV 1393.78 A respectively associated with the formation temperature of log(T / K) = 4.2, 4.0, 4.3, and 4.8. The velocity resolution of IRIS is 
$1 \km\s^{-1}$ (\citealt*{2014SoPh..289.2733D}).

Si\,{\sc IV} shows the characteristics of optically thin line and thus fitted with the single Gaussian while Ni\,{\sc I} is absorption 
line and inverse Gaussian is fitted.
Mg\,{\sc II}\,k and C\,{\sc II}  are fitted with single or double 
Gaussian depending on their profile characteristics.

The corresponding SDO/AIA observations are also taken in the different filters
covering UV/EUV range corresponding to a different temperature range in the solar
atmosphere. AIA provides full-disk multi-wavelength observations of coronal lines
having a spatial resolution of 1.5\arcsec~with a pixel size of 0.6\arcsec~and temporal cadence of 12 s (\citealt{2012SoPh..275...17L}).

The co-aligned Level 2 SDO/AIA data cube has been used in which all the wavelengths are field-of-view matched with 1600 \AA~. To co-align with the raster images of Si IV (1393.78\AA~),  the near-time 171 \AA~ image properly cross-correlated with 1600 \AA~ has been used to compensate for the different resolution of two instruments. However, in our paper, we study the bulk plasma flows from the chosen moss region (in various boxes) by deducing the integrated spectral line-profiles of various IRIS lines. The plage regions are identified in SDO/AIA image data, and the location is mapped onto comparatively high-resolution IRIS data. 

 
\begin{figure*}
\includegraphics[trim=2.5cm 0.5cm 0.2cm 0.2cm,scale=0.6,angle=90]{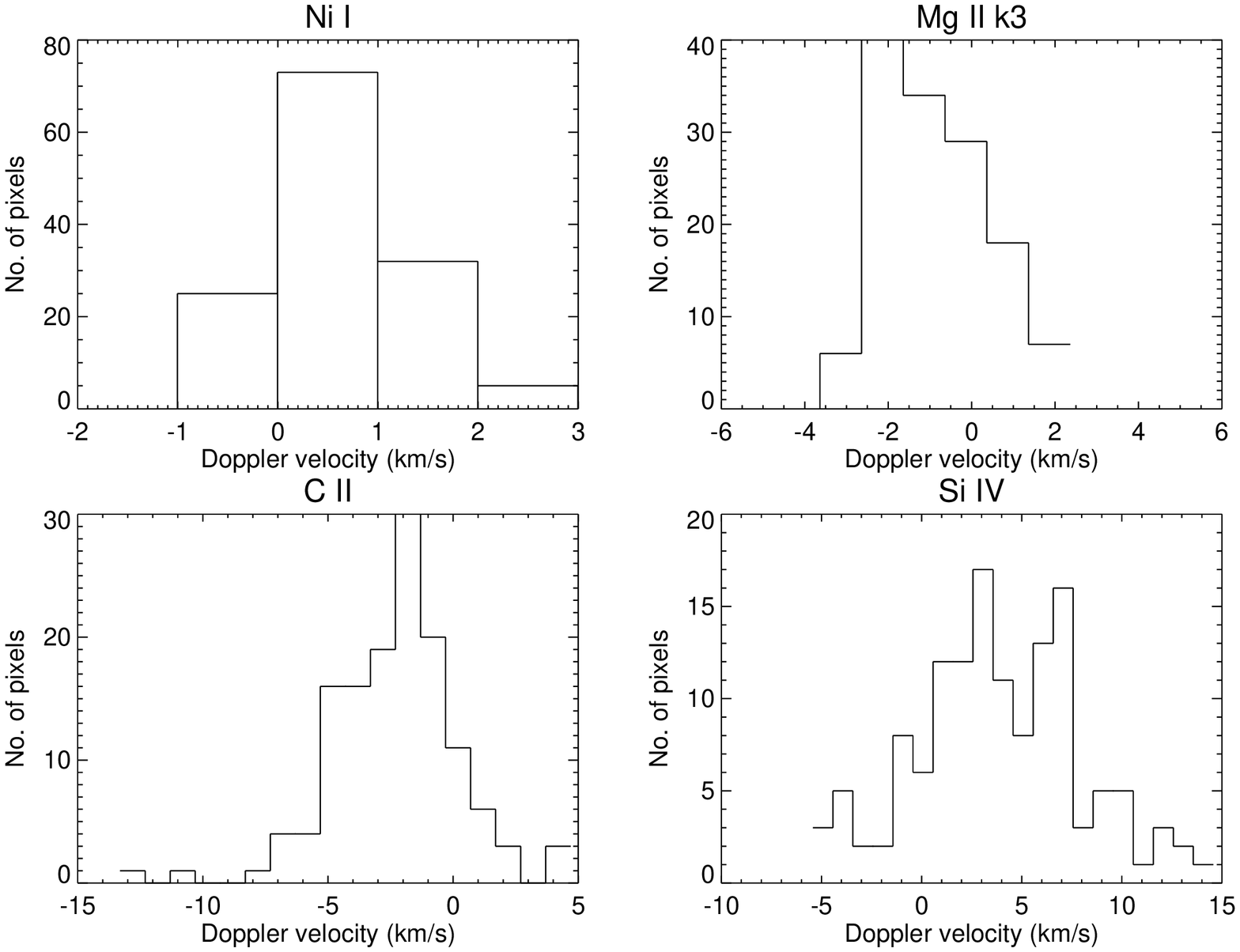}
\vspace{1cm}
\caption{The velocity distributions for different spectral lines corresponding to different temperatures at box B2.} 
\label{fig4}
\end{figure*}

\begin{figure*}
\includegraphics[trim=2.5cm 0.5cm 0.2cm 0.2cm,scale=0.6,angle=90]{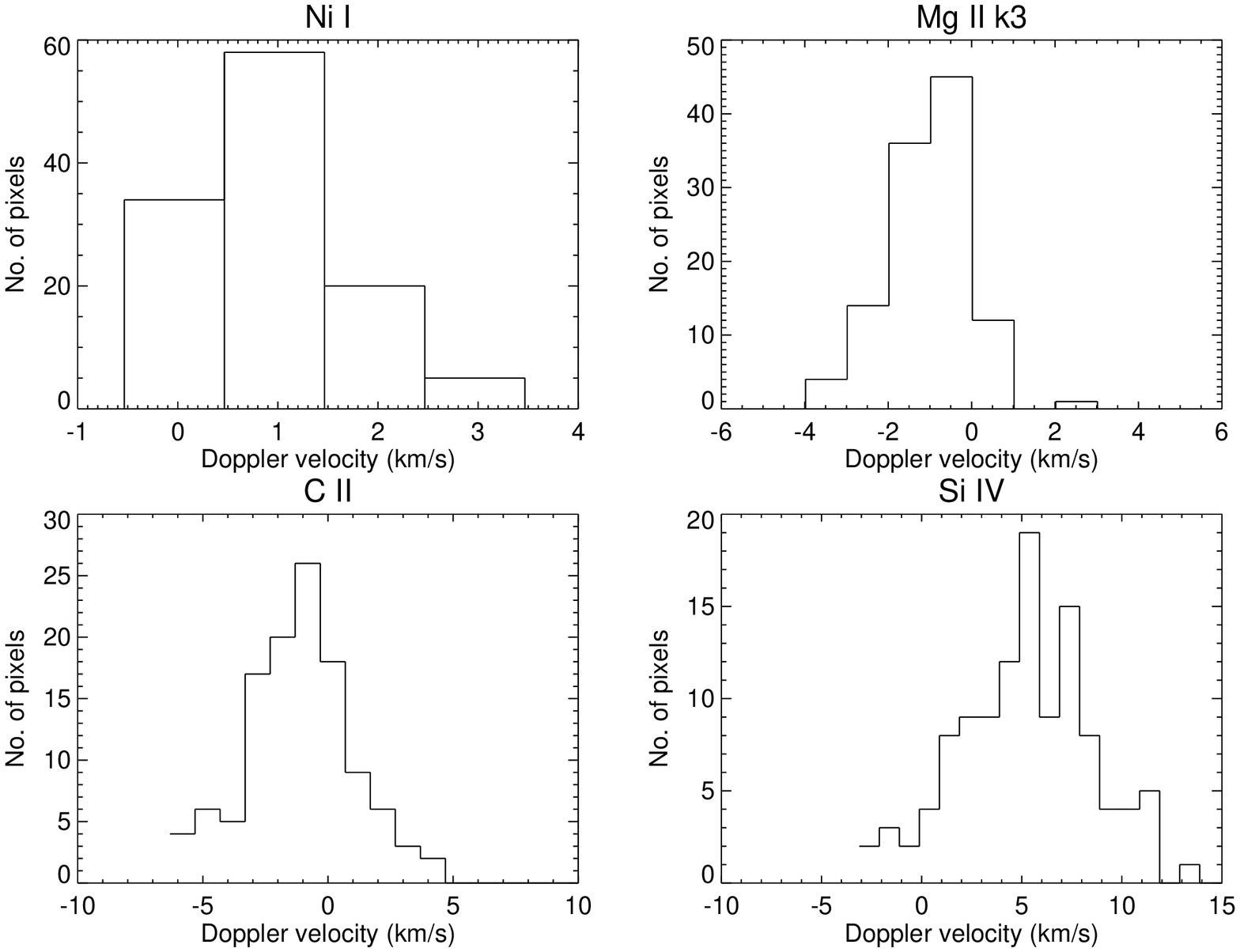}
\vspace{1cm}
\caption{The velocity distributions for different spectral lines corresponding to different temperatures at box B3.}
\label{fig5}
\end{figure*}


\section{Observational Results}

Fig.~\ref{fig0} shows the intensity emission of AR 12529 having plage region and various loops anchored in the moss region are visible in 171 \AA~wavelength of SDO/AIA.
The green emission predominantly indicates the highest emission representing the part of moss region. The yellow box 
is overlaid to show the region of interest (ROI).   

In Fig.~\ref{fig1}, the moss region has been identified with the brightest emission 
in SDO/AIA 193~{\AA} filter. 
The intensity threshold of above 3000 counts (see Fig.~\ref{fig0}) having values double that of plage region surrounding it has been set which is shown by contours 
overlying on the different filters corresponding to different temperature 
ranges from the upper photosphere to corona.
The northern segment of the moss regions where the quiescent coronal loops are 
anchored have been further taken to analyse the flows at the footpoints of these loop systems.
The different bands of AIA show different morphological characteristics of an AR. 
1600~\AA~and 1700~\AA~represent the continuum emission at the upper photosphere showing plage region
near the active region. 
The 304~\AA~gives the chromospheric emission of the plage region. The hot loop structures are 
not distinctly visible. The middle row shows the inner coronal channels where quiescent coronal loops are 
clearly visible, and moss region is identified.  
The last row shows the hot channels of SDO/AIA (335~\AA, 211\AA
, and 94~\AA) where the quiescent loops
taken for our analysis are not visible since quiescent loops are dominated by emissions from the temperatures ranging from 0.7 to 1 MK corresponding to SDO/AIA filters. 
   
The left panel of Fig.~\ref{fig2} shows the HMI map of the region of interest (ROI) indicating the magnetic polarities at the moss region as well as footpoints of the quiescent coronal loops are shown.
The right panel of Fig.~\ref{fig2} is the emission of 193~\AA~line plotted in reverse color 
to identify the footpoints. 
The different boxes of different sizes are then chosen around the footpoints to cover the full strand 
of loop.

Fig.~\ref{fig2c} shows the parametric plots of the Si\,{\sc IV} (1393.7604~\AA~) showing the intensity, Doppler velocity, and FWHM maps where the values has been indicated by the colorbars over the plots. The Doppler velocity maps shows that the TR is dominated by red-shifts even in the plage region surrounding the moss in which footpoints  have been taken for our analysis.

Fig.~\ref{fig2b} shows the DEM maps of the ROI derived by using automated method discussed by \citet{2012A&A...539A.146H}  in which plasma emission at different temperatures has been shown. The colorbar indicates the range of the DEM values. These maps shows the presence of multi-thermal plasma at the footpoints of the quiescent coronal loop systems shown by the boxes around it.


\begin{figure*}
\includegraphics[trim=2.5cm 0.5cm 0.2cm 0.2cm,scale=0.6,angle=90]{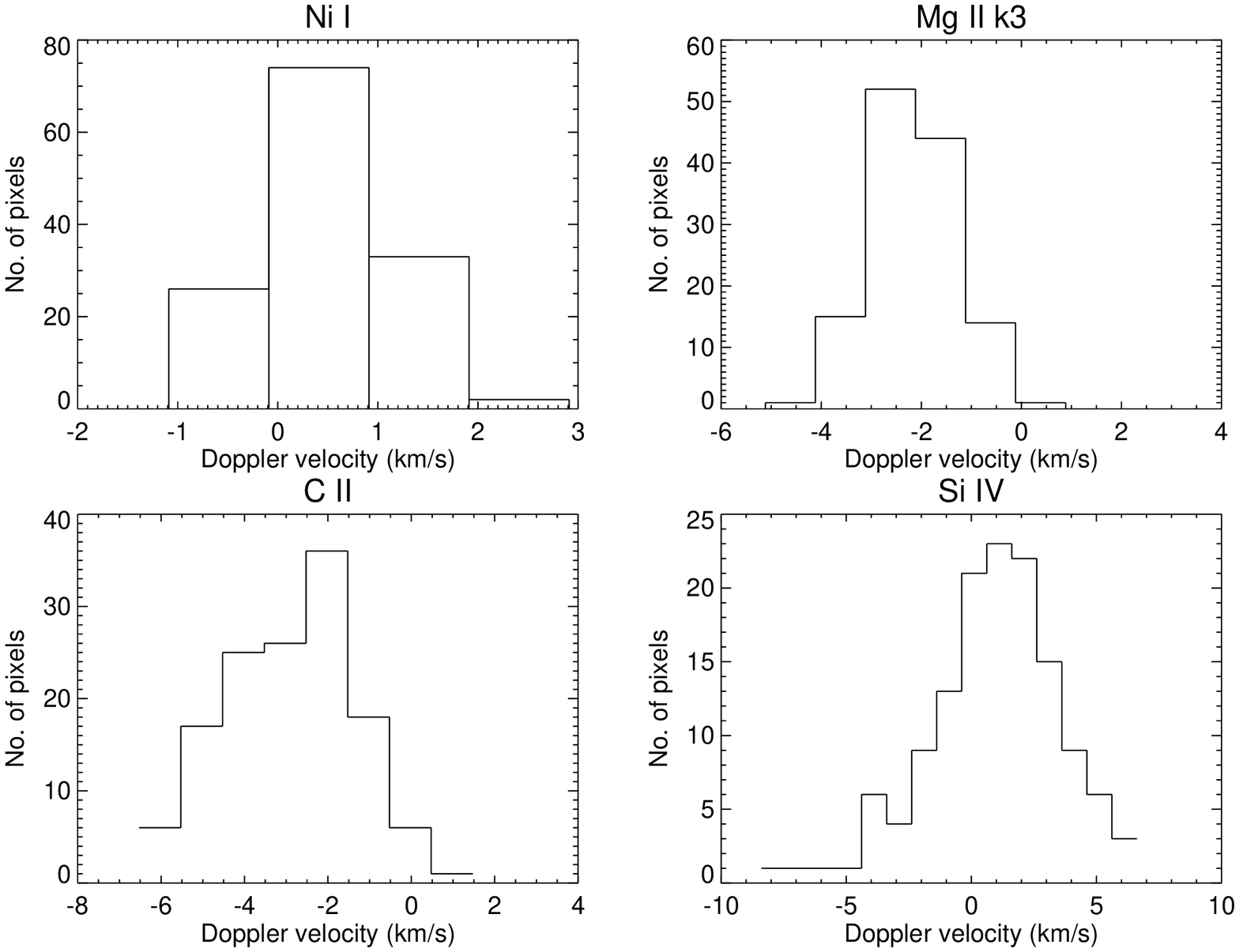}
\vspace{1cm}
\caption{The velocity distributions for different spectral lines corresponding to different temperatures at box B4.}
\label{fig6}
\end{figure*}

\begin{figure*}
\includegraphics[trim=2.5cm 0.5cm 0.2cm 0.2cm,scale=0.6,angle=90]{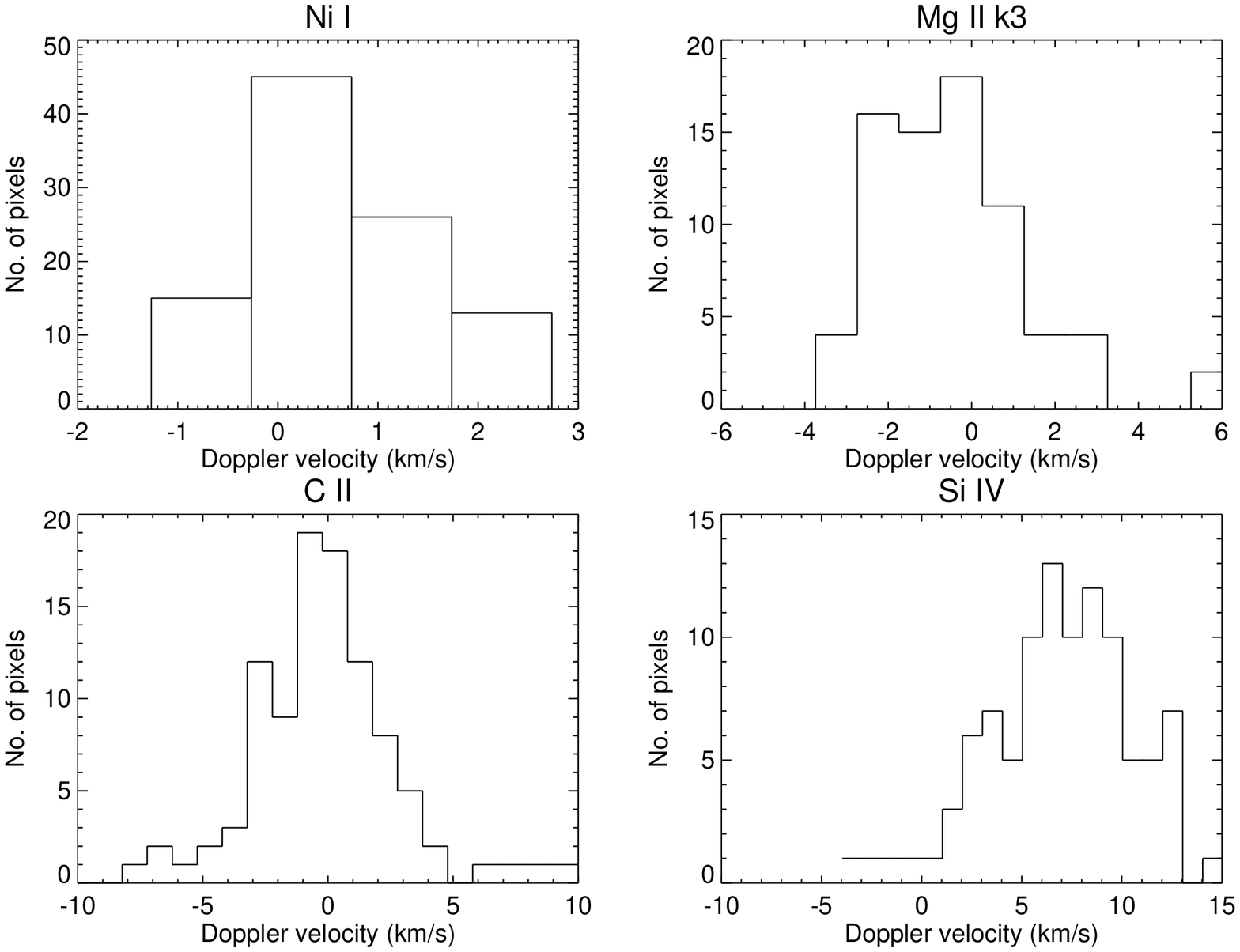}
\vspace{1cm}
\caption{The velocity distributions for different spectral lines corresponding to different temperatures at box B5.}
\label{fig7}
\end{figure*}


The Doppler velocity distribution is thus explored at different locations labelled as B1, B2, B3, B4, and B5.
Positive values (red-shifts) represent downflows, while the negative values (blue-shifts) indicate upflows.
The Doppler velocity at each pixel in first box (B1) for different spectral lines is then shown
Fig.~\ref{fig3}. The velocity distribution for Ni I shows the spread around 0 $\km\s^{-1}$ ranging from ($-0.8~{\rm to}~+2.2) \km\s^{-1}$ .
Mg\,{\sc II}\,k shows the velocities ranging from ($-5~{\rm to}~+1) \km\s^{-1}$  while C\,{\sc II} has ($-5~{\rm to}~+5) \km\s^{-1}$ .
Si\,{\sc IV} shows red-shifts having Doppler velocities ranging from ($0~{\rm to}~+15) \km\s^{-1}$ . 
The histogram of the Doppler velocity for different spectral lines indicate the redshifts in the Si\,{\sc IV} line and 
very small or negligible flows at Ni\,{\sc I}, Mg\,{\sc II}\,k, and C\,{\sc II}.

Similarly, such Doppler velocity distribution is shown in Fig.~\ref{fig4}, 
Fig.~\ref{fig5}, 
Fig.~\ref{fig6}, and 
Fig.~\ref{fig7} for different boxes labelled as B2, B3, B4, and B5.

\begin{figure*}
\includegraphics[trim=2.5cm 0.5cm 0.2cm 0.2cm,scale=0.6,angle=90]{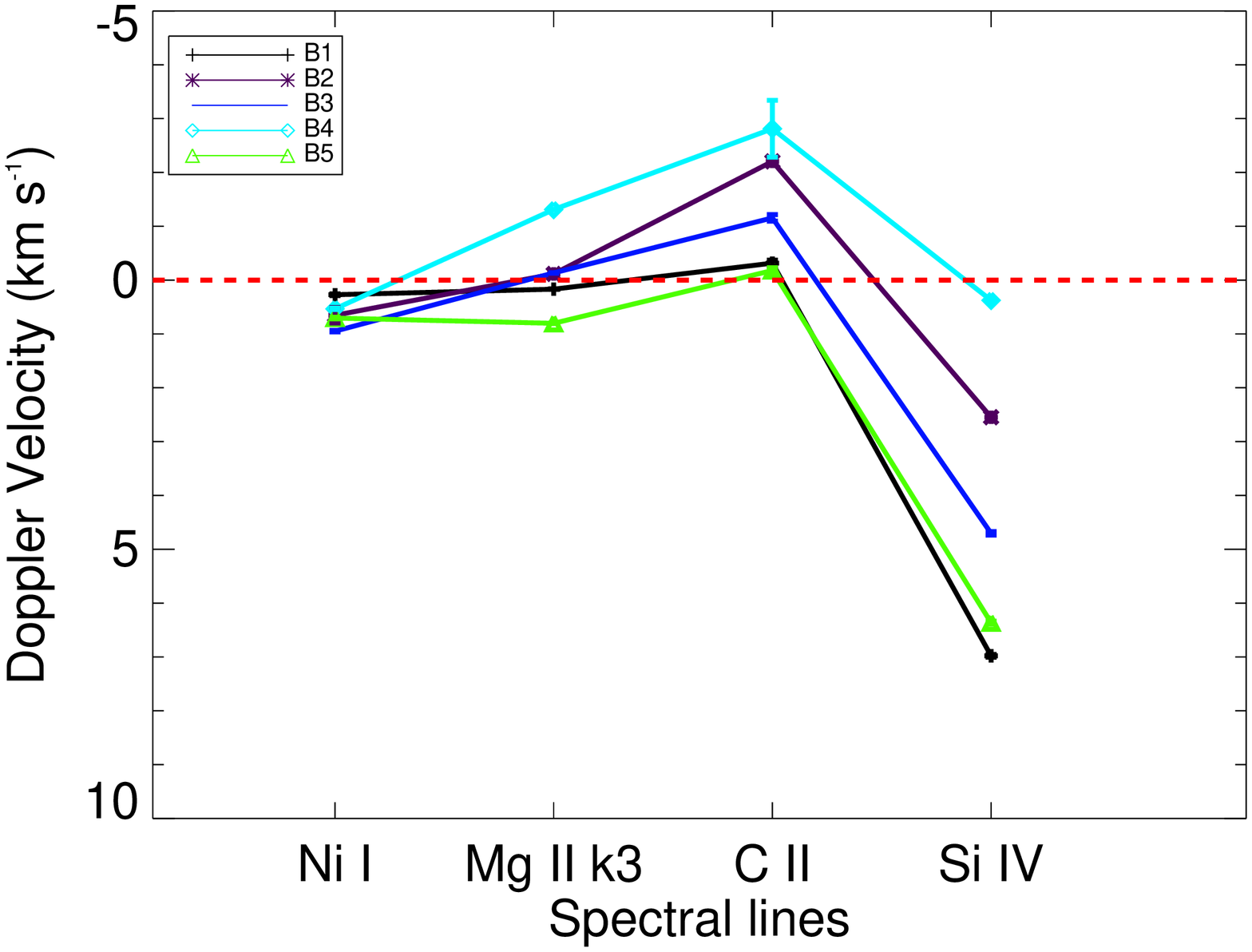}
\vspace{1cm}
\caption{Average Doppler velocity variations for different spectral lines dominating at different heights in the solar atmosphere for boxes B1, B2, b3, B4, and B5 at the footpoints of quiescent coronal loops.}
\label{fig8}
\end{figure*}

Fig.~\ref{fig8} shows the average Doppler shift of different spectral lines
 as a function of their temperatures for different boxes
chosen at the footpoints of the loops. 
Ni I (2799.47 A) corresponds to upper photosphere,
 Mg II k gives emission ranging from mid-chromosphere to upper chromosphere. 
The core defined by (k3) forms ittle higher than the wings at 200 km below TR 
(\citet{2013ApJ...772...90L}). 
C II core gives emission from 2.1 Mm while Si IV corresponds to the TR emission 
(\citet{2015ApJ...811...81R}).

The Doppler velocity of the Ni\,{\sc I} line has negligible values 
indicating almost no flows ($0.27~{\rm to}~0.70) \km\s^{-1}$  corresponding to the photospheric region.
The blueshifts (upflows) show small increment for B2, B4, and B4 ($-0.11~{\rm to}~-0.31) \km\s^{-1}$  while it remains almost same
for B1 (0.16 $\km\s^{-1}$) and B5 ($0.80 \km\s^{-1}$) up to the formation temperature of Mg\,{\sc II}\,k. C\,{\sc II} line shows considerable
blueshifts (upflows) ($-0.17~{\rm to}~-2.81) \km\s^{-1}$ but it is still negligible as compared to chromospheric flows.
The Doppler velocity variation at Si IV shows prevalent redshifts (downflows) at all the locations corresponding to TR flows ($0.37~{\rm to}~6.97) \km\s^{-1}$. 
The 1-sigma error is shown as error bars which is difficult to visualize 
in Fig.~\ref{fig8} owing to its very small values.


\section{Discussions and Conclusions}  

The co-spatial multi-spectral Doppler velocity trend at the footpoints of quiescent coronal loops has been studied. The Doppler velocity variation shows small flows (upflows/downflows) for Ni\,{\sc I} and Mg\,{\sc II}\,k the photospheric as well as the chromospheric region. C\,{\sc II} shows very blue-shifts  ($-0.1~{\rm to}~-2.81) \km\s^{-1}$) indicating small upflows at upper chromospheric region.
The Doppler velocities then change to red-shifts at the formation temperature of Si\,{\sc IV} line corresponding to the TR.

It has been previously shown that the moss regions show significant red-shifts (downflows) 
in the TR explaining the low-frequency heating (\citealt{2010ApJ...710L..39B}). 
The high- and low- frequency mechanisms depend upon the time taken by the loops to cool down as compared to 
heating frequency (\citealt{2008A&A...481L..53T}).

Our study of the flows at the quiescent coronal loops shows similar characteristics as the dynamically active loops though the velocity values are less. 
The plasma predominantly shows red-shifts at TR temperatures which corroborates with the low-frequency heating of 
loops in the coronal part of the solar atmosphere.
These observations thus agree with the coronal loops heated up by 
low-frequency nanoflares via impulsive heating mechanism.
Also, \citealt{2006ApJ...647.1452P} observed the symmetric profiles for steady heating in the loops.
Though it is possible to have asymmetries in the individual profile for which velocity distribution has been observed, our speculation supports the nano-flare driven impulsive heating mechanism for the quiescent coronal loops.

The asymmetries may also cause these Doppler variation in the spectral profiles due to a difference in the pressures (Mariska \& Boris 1983). So, other possibilities cannot be ruled out. (\citealt{1983ApJ...267..409M}). So, other possibilities cannot be ruled out.

\begin{acknowledgements}
One of us (Yamini K. Rao) is fully supported by the financial grant from the ISRO RESPOND project. 
We acknowledge the use of IRIS observations. IRIS is a NASA small explorer mission developed and operated by Lockheed Martin Solar and Astrophysics Laboratory (LMSAL) with mission operations executed at NASA Ames Research Center and major contributions to downlink communications funded by the Norwegian Space Center (NSC, Norway) through an ESA PRODEX contract.
\end{acknowledgements}




%

\end{document}